\documentclass[journal]{IEEEtran}
\usepackage{amsmath,amssymb,amsfonts}
\usepackage{algorithmic}
\usepackage{graphicx}
\usepackage{textcomp}
\usepackage{xcolor}
\usepackage[hidelinks]{hyperref}
\usepackage[sorting=none]{biblatex}
\begin{document}

\title{Analysis and Modeling of the Hybrid Vessel's Electrical Power System}

\author{Matthijs Mosselaar, Zoran Malba\v{s}i\'{c}, Aihui Fu and Aleksandra Leki\'{c}% <-this % stops a space
\thanks{M. Mosselaar, A. Fu and A. Leki\'{c} are with the Faculty of Electrical Engineering, Mathematics and Computer Science, Delft University of Technology, Delft, The Netherlands (correspondence e-mails: \{A.Fu, A.Lekic\}@tudelft.nl) \\
Z. Malba\v{s}i\'{c} is with Power conversion \& hybrid solutions, Alewijnse Netherlands, The Netherlands, \href{mailto:z.malbasic@alewijnse.com}{Z.Malbasic@alewijnse.com}.}% <-this % stops a space
}

\maketitle

\begin{abstract}
With the maritime industry poised on the cusp of a hybrid revolution, the design and analysis of advanced vessel systems have become paramount for engineers. This paper presents AC and DC electrical hybrid power system models in ETAP, the simulation software that can be adapted to engineer future hybrid vessels. These models are also a step towards a digital twin model that can help in troubleshooting and preventing issues, reducing risk and engineering time. The testing of the models is focused on time domain analysis, short-circuit currents, and protection \& coordination. The models are based on actual vessels and manufacturer parameters are used where available.
\end{abstract}\medskip


\section{Introduction}
The maritime industry is undergoing a transformative shift toward decarbonization. Spearheading this movement is the Getting to Zero Coalition, which is an alliance of over 200 organizations within the maritime, energy, infrastructure, and finance sectors. Their goal is to get commercially viable deep-sea zero-emission vessels powered by zero-emission fuels into operation by 2030 for a path toward full decarbonization by 2050 \cite{global-maritime-forum}.  Hybrid ships and vessels play a key role in this transition and it is important to have a comprehensive understanding of hybrid power systems imperative.

Traditionally, maritime power systems are low voltage, high power systems that are built up from a main busbar to which the main generators and the heavy loads are connected. The voltage level of the systems is usually 690 V or 400 V and the main busbar can often be bifurcated into port and starboard side busbars, which are connectable. From the main busbar, the power can be further distributed on a lower voltage level of 400 V or 230 V with frequency standards set at either 50 Hz or 60 Hz. Many systems also accommodate an emergency generator, which can be linked to the main busbar directly or via another intermediary busbar.

The allure of hybrid systems isn't confined to just new vessel constructions. Existing vessels can be retrofitted with batteries, commonly undertaken during mid-life refits. A refit can include repairing, fixing, restoring, renewing, mending, and renovating the vessel, depending on its needs. As long as there is space on the vessel to fit the battery and its converter, it should be possible to convert a conventional diesel-powered vessel to a hybrid vessel. For vessels equipped with an AC main busbar, battery integration is achieved using a bidirectional inverter. Conversely, vessels with a DC main busbar can connect the battery directly or through a DC/DC converter, akin to other DC components.

This paper uses the ETAP simulation software (version 22.5.0) to design two adaptable  electrical hybrid power system models that can be adapted for the engineering of future hybrid vessels. The first model is based on an AC main busbar and the second system is based on a DC main busbar. Authentic vessels inform both models, preferring manufacturer-specific parameters wherever feasible. This paper focuses on time domain analysis, short-circuit current calculations, and protection \& coordination. 
%\vspace{20pt}

\section{Hybrid AC grid simulation}
\begin{figure}[htbp]
\centerline{\includegraphics[width=0.5\textwidth]{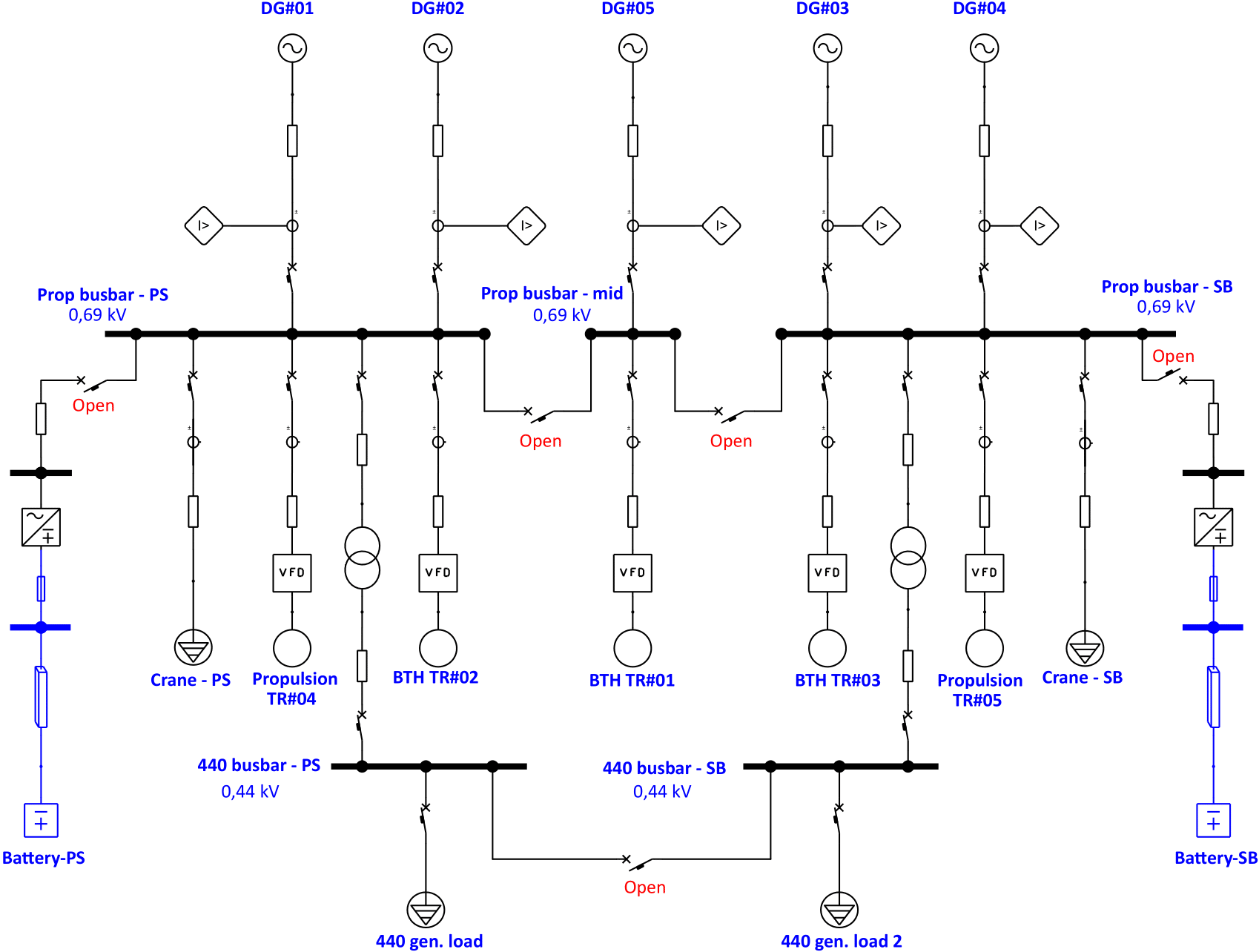}}
\caption{Single line diagram of the studied AC grid}
\label{fig:ac-sld}
\end{figure}

The first system is based on an AC bus and is shown in figure \ref{fig:ac-sld}. It is the single-line diagram of a cable-laying vessel, and the system can be divided into three parts according to the three different sections of the propulsion busbar. Both the port side (PS) and starboard side (SB) sections mirror each other in design. They are each anchored by two 690 V generators of disparate capacities, which serve as the principal power sources for the system. The detailed parameters of the generators can be found in table \ref{tab:gen-parameters}. The two inverters connecting the 1500 kWh batteries are rated at 1500 kW which limits the maximum charge and discharge rate of the batteries at 1 C.  The system comprises three bow thrusters and two propulsion thrusters, rated at 1000kW and 2100kW respectively, both with a power factor of 0.85. The cranes are modeled as lumped loads with a 20 \% static part and an 80 \% motor part with a 0.75 power factor. This configuration culminates in a cumulative load of 746.7 kVA. The 440 generation loads are also modeled as lumped loads but with a 65 \% static and 35 \% motor part at 0.90 power factor.
\vspace{-10pt}

\begin{table}[htbp]
\centering
\caption{Generator parameters from datasheet}
\begin{tabular}{l|c|c|c|}
\cline{2-4}
 & \multicolumn{1}{l|}{\textbf{\begin{tabular}[c]{@{}l@{}}DG\#01\\ DG\#04\end{tabular}}} & \multicolumn{1}{l|}{\textbf{\begin{tabular}[c]{@{}l@{}}DG\#02\\ DG\#03\end{tabular}}} & \multicolumn{1}{l|}{\textbf{DG\#05}} \\ \hline
\multicolumn{1}{|l|}{Rated power (kVA)} & 2395 & 3213 & 1713 \\ \hline
\multicolumn{1}{|l|}{Rated power (kW)} & 1916 & 2570 & 1370 \\ \hline
\multicolumn{1}{|l|}{Voltage (V)} & 690 & 690 & 690 \\ \hline
\multicolumn{1}{|l|}{Current (A)} & 2004 & 2688 & 1433 \\ \hline
\multicolumn{1}{|l|}{Frequency (Hz)} & 60 & 60 & 60 \\ \hline
\multicolumn{1}{|l|}{Power factor} & 0.80 & 0.80 & 0.80 \\ \hline
\multicolumn{1}{|l|}{Speed (rpm)} & 720 & 720 & 1800 \\ \hline
\multicolumn{1}{|l|}{\begin{tabular}[c]{@{}l@{}}Winding resistance\\ per phase ($m\Omega$)\end{tabular}} & 1.02 & 0.70 & 1.46 \\ \hline
\end{tabular}
\label{tab:gen-parameters}
\end{table}

\subsection{Time domain analysis}
Time domain simulations within maritime contexts often spotlight two salient scenarios: peak shaving and Dynamic Positioning (DP) mode. Within these scenarios, the battery operates analogously to a virtual generator. In the peak shaving study case, the goal is to demonstrate the inverter's capacity for peak shaving. This is achieved through an ETAP User-Defined Model (UDM) for the inverter, which continually monitors the power of the generators throughout the simulation. If the power were to rise above 80 \% of the generator's capability, the inverter is told to supply the difference, keeping the loading steady at 80 \%. For DG\#01, 80 \% means 1.5 MW and for DG\#02, 80 \% means 2.0 MW. The output power of the inverter is limited according to its rating. The inverter is also set to provide reactive power if it were to go above 1 Mvar or 1.5 Mvar for generator DG\#01 and DG\#02 respectively. The reactive power ratings are chosen arbitrarily.
\vspace{-3pt}
\begin{figure}[htbp]
\centerline{\includegraphics[width=0.5\textwidth]{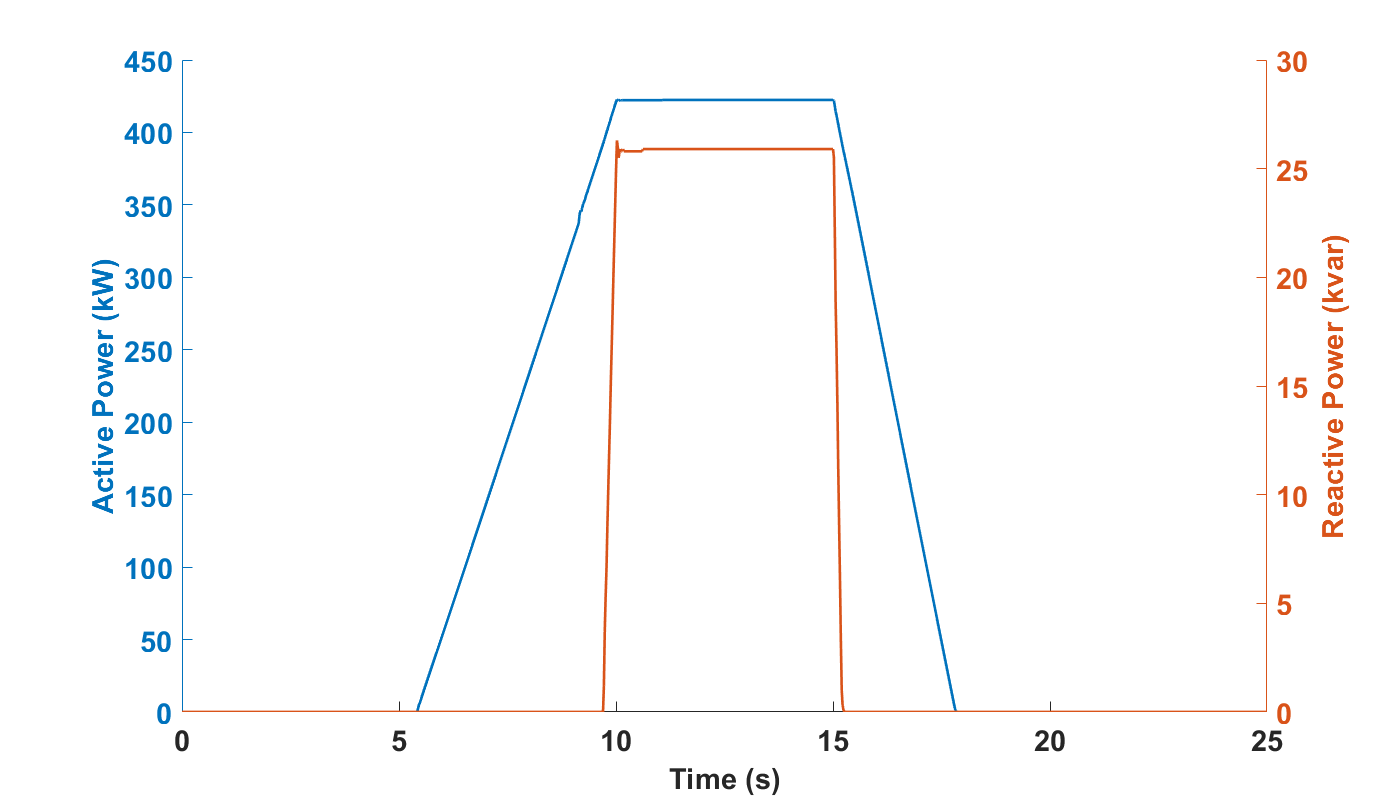}}
\caption{Inverter active and reactive power}
\label{fig:PS-inv-kW-kvar}
\end{figure}

\begin{figure}[htbp]
\centerline{\includegraphics[width=0.5\textwidth]{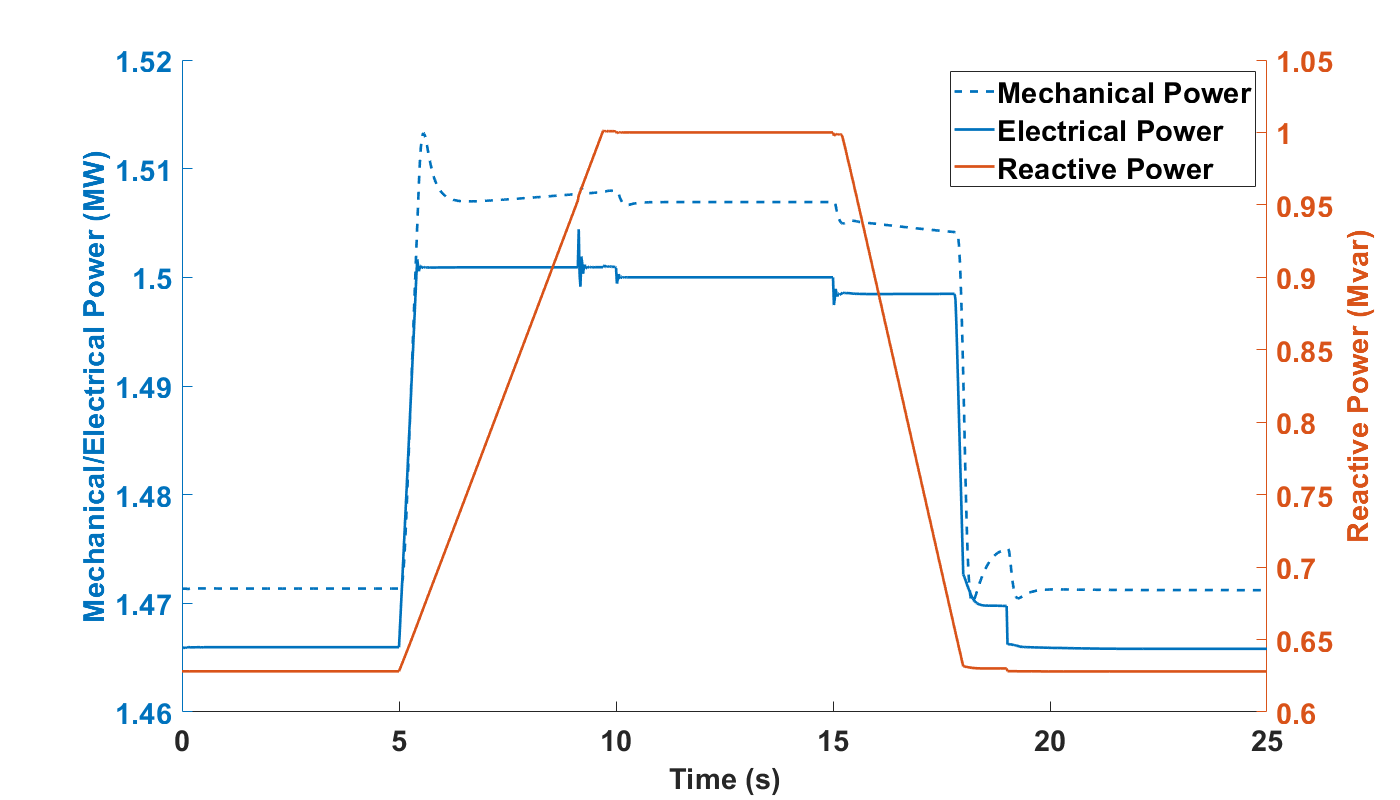}}
\caption{Generator active and reactive powers}
\label{fig:PS-gen-MW-MW-Mvar}
\end{figure}

For the simulation, only generator DG\#01 is active and bears a load slightly below 1.5 MW. At $t = 5$s, a lumped load is increased and later decreased, taking the loading of the generator above the 1.5 MW threshold. Figure \ref{fig:PS-inv-kW-kvar} captures the inverter's response. Slightly after $t = 5$s, when the generator power has reached 1.5 MW, the inverter starts to provide power according to the increasing load. Only at the very end of the load ramp does the reactive power of the generator go above 1 Mvar. Figure \ref{fig:PS-gen-MW-MW-Mvar} shows the mechanical power, electrical power, and reactive power of the connected generator. As can be seen, the active power of the generator does not go above 1.5 MW and the reactive power does not go above 1 Mvar.

The DP-mode simulation envisions a sudden generator disconnect, simulated by interrupting the circuit breaker connecting the generator to the main busbar. The inverter is then set to supply the power as the now-disconnected generator, but within its rated limits. This is again done through a UDM that monitors the power of the generators and stores a delayed value, such that it can provide that value in case of a sudden loss of power.

\begin{figure}[htbp]
\centerline{\includegraphics[width=0.5\textwidth]{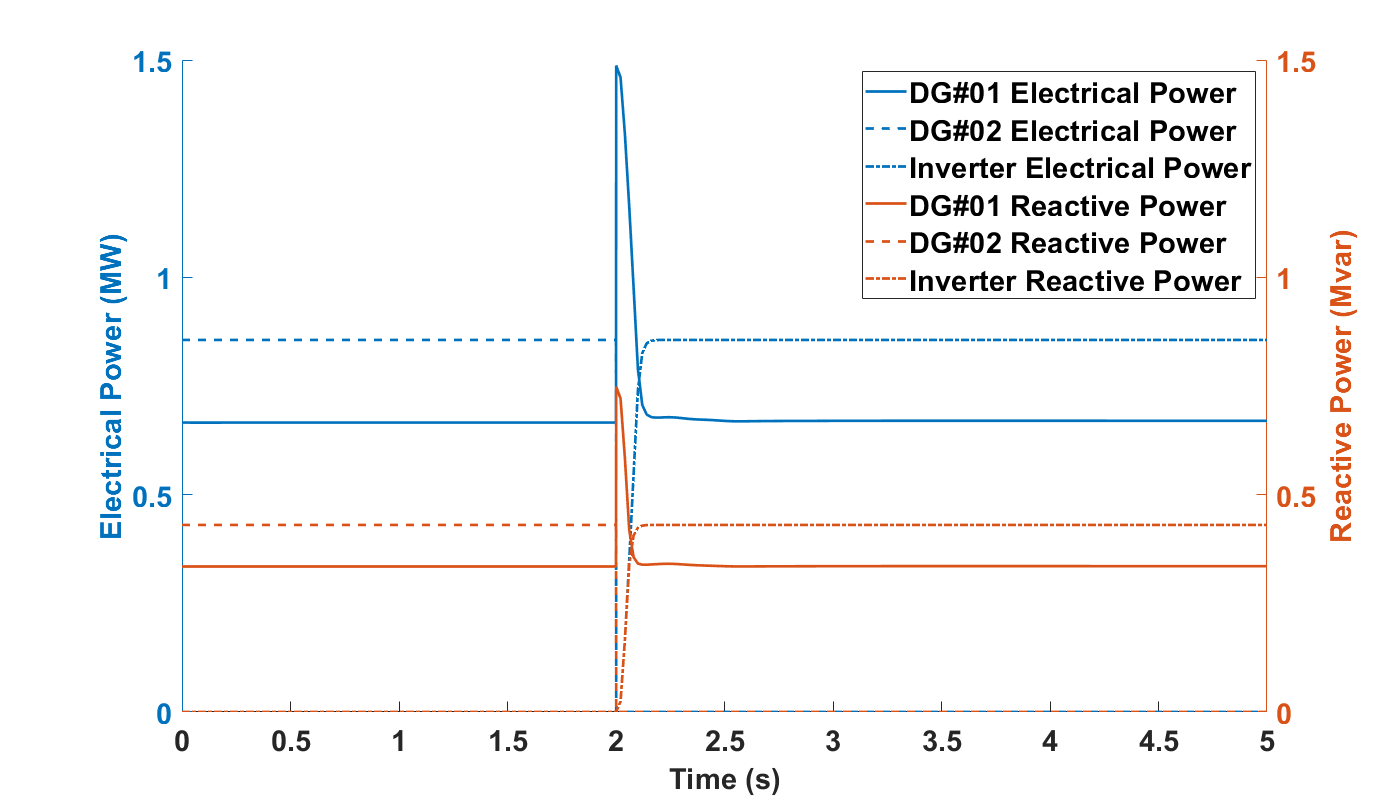}}
\caption{Inverter and generators active and reactive power}
\label{fig:DP-inv-gens-kw-kvar}
\end{figure}

In this simulation, both DG\#01 and DG\#02 are connected, and DG\#02 will suddenly disconnect. Figure \ref{fig:DP-inv-gens-kw-kvar} shows the active and reactive power of the inverter and generators. The moment DG\#02 disconnects at $t = 2$s, the inverter starts providing power. After a short spike in the power request from the unaffected generator, the generator power level returns to the original level.

\vspace{-3pt}

The Critical Clearing Time (CCT) has also been analyzed using ETAP. For all generators, the applied fault is a worst-case fault where the generators are loaded at 90 \%, and the fault is a short-line fault at only 1 \% cable distance. The CCT of generator DG\#01 and DG\#04 is 542 ms. The CCT of generator DG\#02 and DG\#03 is 564 ms, and the CCT of generator DG\#05 is 574 ms.

\subsection{Short-circuit calculations}
For short-circuit calculations in the maritime industry, we use the IEC 61363 \cite{iec-61363-1-1998} standard. Results for the calculated generator short-circuit currents are shown at the top of table \ref{tab:sc-gens}. The results of the simulated short-circuit currents using the IEC 61363 calculation tool in ETAP are shown at the bottom of the same table.

\vspace{-10pt}
\begin{table}[htbp]
\centering
\caption{Calculated (top) and simulated (bottom) generator short-circuit currents}
\begin{tabular}{l|l|l|l|}
\cline{2-4}
 & \textbf{\begin{tabular}[c]{@{}l@{}}DG\#01\\ DG\#04\end{tabular}} & \textbf{\begin{tabular}[c]{@{}l@{}}DG\#02\\ DG\#03\end{tabular}} & \textbf{DG\#05} \\ \hline
\multicolumn{1}{|l|}{$I_{ac}(t)$} & 13.759 kA & 17.971 kA & 12.733 kA \\ \hline
\multicolumn{1}{|l|}{$I_{dc}(t)$} & 16.311 kA & 21.076 kA & 15.458 kA \\ \hline
\multicolumn{1}{|l|}{$I_p(t)$} & 35.769 kA & 46.490 kA & 33.521 kA \\ \hline
 &  &  &  \\ \hline
\multicolumn{1}{|l|}{Iac(t)} & 11.231 kA & 14.710 kA & 11.662 kA \\ \hline
\multicolumn{1}{|l|}{idc(t)} & 16.265 kA & 21.264 kA & 14.867 kA \\ \hline
\multicolumn{1}{|l|}{ip(t)} & 32.148 kA & 42.067 kA & 31.359 kA \\ \hline
\end{tabular}
\label{tab:sc-gens}
\end{table}
\vspace{-3pt}

All values used for the theoretical calculation are taken from the datasheets, just like the values for the simulation. The short-circuit currents $i_p$ are a sum of the AC $I_{ac}$ and DC $i_{dc}$ components. When comparing the top and bottom values of table \ref{tab:sc-gens}, there is a distinct difference between the two. One possible explanation for this difference could be that ETAP uses open-circuit time constants in its simulation model and short-circuit time constants are used for the theoretical calculation. The datasheets of generators DG\#01 and DG\#02 only provide short-circuit time constants and their time constants have been converted for the simulation model. The datasheet of generator DG\#05 provides both open-circuit and short-circuit time constants and its simulated short-circuit current is closer to the theoretical value compared to the other two generators. This means a conversion error could be a contributing factor to the difference seen between theoretical and simulated results, but a much larger sample size of generators is needed to prove this.

A second explanation for the difference can be found in the AC part of the short-circuit calculation. Looking at the \emph{$I_{ac}(t)$} term specifically, there is the term $I_{kd}$ which represents the steady-state short-circuit current which is generally obtained from the manufacturer. The datasheet value of this current is used in the theoretical calculation, but ETAP calculates this value based on the system impedance at the time of the fault.

A further contribution to the short-circuit current comes from the variable frequency drives and the two lumped loads in the simulation. Specific contribution from the Variable Frequency Drives is not available, but the short-circuit contribution is typically set at 150 \% of the nominal current. This same short-circuit contribution is also used in the simulation software. The cranes are modeled as lumped loads with a 20 \% static part and 80 \% motor part. The motor part of the lumped load will contribute to the short-circuit current. ETAP shows a contribution of 4.831 kA and using the IEC 61363 with the same X/R ratio, a contribution of 4.488 kA is calculated.

\subsection{Protection \& coordination}
In the protection and coordination study, it is made sure that the system is able to selectively trip circuit breakers to protect its components during a fault condition. Important aspects of this study are the decrement curves of the generators and the Time Current Curves of the circuit breakers. These are acquired directly from the manufacturer.

In case of a short-circuit fault between the generator and the generator circuit breaker, the circuit breaker must be opened as soon as possible to prevent damage to the generator and instability to the system as discussed in the previous section. However, it is also important that only the generator circuit breaker switches off. If the breaker detects a fault, but it's somewhere else in the system, the breaker should wait for another breaker to clear the fault. If another breaker does not clear the fault within a specific time, the breaker should still open.

In practice, this means that if a fault occurs between generator DG\#01 and its breaker, the breaker will see a fault current in the direction of the generator. The breaker will then send a lock signal to the other breakers that are contributing to the fault. This will start a delay where, if the fault is not cleared within this time, the breakers that received the lock signal will open anyway. In the case where the three propulsion busbar sections are connected, if the breaker connecting the port side and the mid busbar receives a lock signal, it will pass on the lock signal to the other active breakers that are contributing to the fault current.

ETAP's Sequence-of-Operation cannot implement lock signals, but it does verify that all breakers detect the fault and can switch off quickly. Without implementing the lock signals, ETAP shows that all circuit breakers detecting the fault will open after 216 ms, which is within the critical clearing time of all generators. This also shows why no instantaneous tripping is implemented, but only (directional) short- and long-term. Since the fault current will depend on the exact location and the loading of the generators, it would be very difficult to set a current threshold for all breakers for all situations. The protection systems need a bit of time to send out lock signals to maintain selectivity and prevent a black-out.

\section{Hybrid DC grid simulation}
\begin{figure}[htbp]
\centerline{\includegraphics[width=0.5\textwidth]{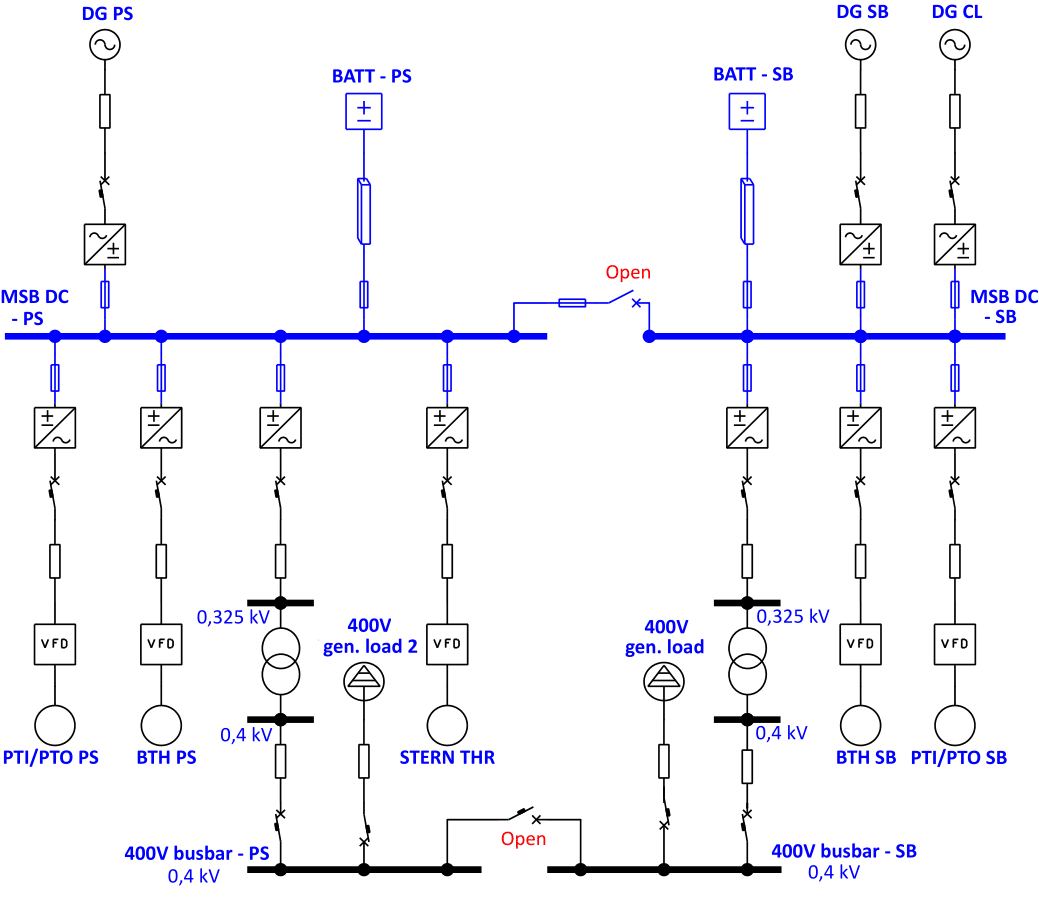}}
\caption{Single line diagram of the studied DC grid}
\label{fig:dc-sld}
\end{figure}

The second system is based on a DC grid and is shown in figure \ref{fig:dc-sld}. It is the single-line diagram of a superyacht. This system closely mirrors the previously described AC grid in numerous aspects; however, there are several salient differences worth noting. In this system, the main busbar is divided into only two sections. Unlike the AC grid, the port and starboard sides of this grid are not symmetrical. Three same-sized generators are present, with just one on the port side. Besides this single generator, the port side comprises a battery, a primary propulsion thruster, a pair of smaller thrusters, and a linkage to the 400 V AC distribution board. The port side also has the option to connect a shore connection instead of the generator, but this is not modeled.

All three generators shown in figure \ref{fig:dc-sld} are identical, and their parameters are listed in table \ref{tab:spec_dc_gens}. The converters are rated at 850 A and will be able to cover the full load of the generators. The two batteries are directly connected to the busbar without any converters. The two propulsion PTI/PTO thrusters are rated at 550 kW at 0.85 power factor and have an inverter rated at 1150 A. The bow thrusters and stern thruster are equal to 200 kW at 0.85 power factor and have an inverter rated at 460A. The transformers are rated at 600 kVA and have grid inverters rated at 2060 A. Finally, the 400 gen. loads are modeled at 400 kVA and have a 65 \% static and 35 \% motor part at 0.85 power factor.

\begin{table}[htbp]
\centering
\caption{Specification of DC grid generators from datasheet}
\begin{tabular}{l|c|}
\cline{2-2}
 & \multicolumn{1}{l|}{\textbf{Generators}} \\ \hline
\multicolumn{1}{|l|}{Rated power (kVA)} & 582 \\ \hline
\multicolumn{1}{|l|}{Rated power (kW)} & 465.6 \\ \hline
\multicolumn{1}{|l|}{Voltage (V)} & 400 \\ \hline
\multicolumn{1}{|l|}{Current (A)} & 840 \\ \hline
\multicolumn{1}{|l|}{Frequency (Hz)} & 50 \\ \hline
\multicolumn{1}{|l|}{Power factor} & 0.80 \\ \hline
\multicolumn{1}{|l|}{Poles} & 4 \\ \hline
\multicolumn{1}{|l|}{Speed (rpm)} & 1500 \\ \hline
\multicolumn{1}{|l|}{\begin{tabular}[c]{@{}l@{}}Winding resistance\\ per phase ($m\Omega$)\end{tabular}} & 3.4 \\ \hline
\end{tabular}
\label{tab:spec_dc_gens}
\end{table}

\subsection{Time domain analysis}
In the transient stability study, the system is modeled in the time domain, but limitations in ETAP were encountered. It is not possible to properly use the battery in time domain analysis. The integrated BMS can only charge or discharge the battery based on a bus voltage or loading. This is useful for renewable resources but not for studying peak shaving for example. A second limitation is that it is impossible to study the DC part of the grid. The two AC parts (generators and loads) are two completely separate systems and there is no power balance between the two. To 'solve' this, the generators are connected to inverters instead of charger components. This is because the inverters can use a UDM, and the chargers can not. The model is programmed to restore the power balance, and the DC part is ignored. It takes the power that is provided at the loads by the inverters and sets that as the power required for the generator. In the future, more advanced capabilities can be programmed into the UDM, to show power sharing between the generators or even 'including' the battery, but these functionalities would be difficult to showcase in ETAP's current version and thus don't add much value to this paper. ETAP can't directly showcase the parameters and variables used in the UDM.

\subsection{Short-circuit calculation}
The components contributing to the short-circuit current on the DC bus are the battery and the converters. The battery manufacturer gives the short-circuit current that the battery will provide, but how much the capacitors in the converters will contribute is not known. The IEC 61660 \cite{iec-61660-1-1997} is the standard for DC short-circuit calculations, but it dates from 1997.  The standard does not provide any calculations for Lithium-Ion batteries, only for Lead-Acid batteries. A lithium-Ion battery's short circuit current curve is similar to that of a capacitor \cite{sang2016study}, but what if the battery is connected using a DC/DC converter? Currently, no standard clearly describes the contribution of all modern-day components, and educated guesses have to be made instead.

\begin{figure}[htbp]
    \centering
    \includegraphics[width=0.48\textwidth]{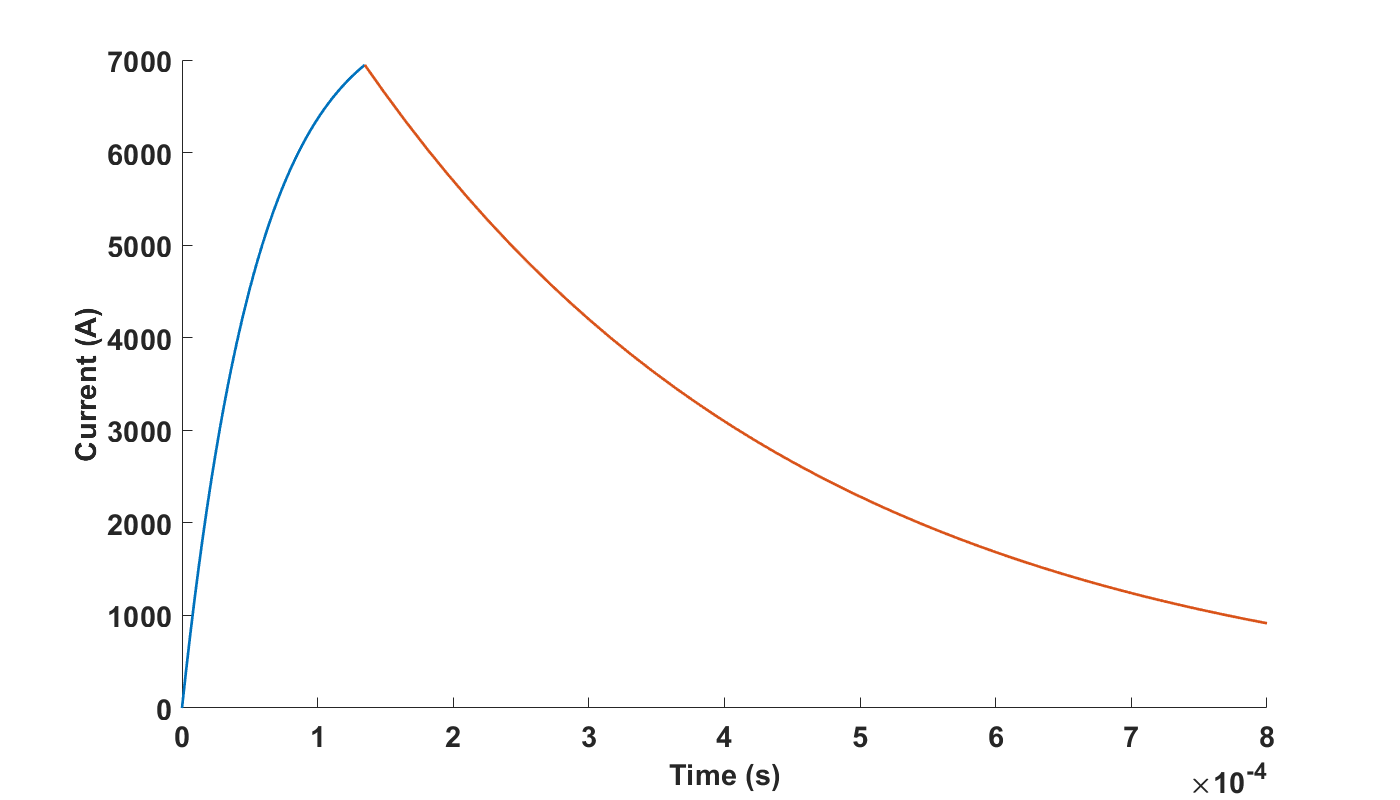}
    \caption{Capacitor short-circuit current in time}
    \label{fig:cap_current}
\end{figure}

The current from a capacitor can be calculated according to the IEC 61660. To get an idea of the peak current and time constant of a capacitor short-circuit, the formulas from the IEC 61660 are used to calculate the short-circuit current of a 2400$\mu$F capacitor with a 54.7m$\Omega$ series resistance and a 5.5$\mu$H series inductance. This graph is shown in figure \ref{fig:cap_current}. The peak current reached is 6.95 kA in 0.134 ms. For comparison, the short-circuit current of the battery is specified as 14.9 kA with an L/R of 0.16 ms.

Similar to the AC grid simulation section, the short-circuit current of the generators can be calculated according to the AC IEC 61363 standard. Then, the short-circuit current is also simulated using ETAP's IEC 61363 calculation tool. The result of both is shown in table \ref{tab:gen-sc-calc-dc}.\medskip
\vspace{-3pt}

Again, there is a slight difference between the calculated and simulated values of the generator short-circuit current. All values that are used for the theoretical calculation are taken from the datasheet, just like the values for the simulation, with one exception. The datasheet did provide an open-circuit transient time constant for the simulation, but not an open-circuit subtransient time constant. This difference between the calculated and simulated value is 5.9 \%, which is actually one percent less than generator DG\#05 from the previous section.

%\vspace{-5pt}
\begin{table}[htbp]
\centering
\caption{Generator short-circuit current}
\begin{tabular}{|l|l|l|l|}
\hline
\textbf{Method} & $\mathbf{I_{ac}(t)}$ & $\mathbf{I_{dc}(t)}$ & $\mathbf{I_{p}(t)}$ \\ \hline
Calculation & 7.035 kA & 7.689 kA & 17.638 kA \\ \hline
Simulation & 6.648 kA & 7.352 kA & 16.648 kA \\ \hline
\end{tabular}
\label{tab:gen-sc-calc-dc}
\end{table}
%\vspace{-3pt}

As mentioned before, the size of the capacitors in the converters is not known and ETAP also does not take them into account. For the charger element, ETAP considers a DC fault current of 150 \% of the full load current, and the inverter elements have no contribution to the DC fault current. It is also impossible to manually add a capacitor to the DC grid as DC capacitors are not included in the current version. This means that, in ETAP, the total fault current, in case of a fault on the port side DC bus, would be the sum of the battery fault current and the 1.275 kA provided by the charger element connecting the generator to the DC bus.

\subsection{Protection \& coordination}
The protection and coordination study is much different for the DC bus system than the AC system. The main challenge for selectivity will be on the 400V AC distribution level. For the DC busbar, finding the right fuse that can sustain the full load of a converter but also disconnect very quickly in case of a short circuit will be challenging. In this system, it will be an advantage that the battery is directly connected to the DC busbar as the short-circuit level of the battery should stay close to the value from the datasheet, regardless of the state of charge. This is because the battery's internal resistance is very low and will only significantly increase at a very low state of charge. Since the battery state of charge is predicted to approach 25 \% as the lowest value, this should not be an issue.

The fuses aim to disconnect the fault current before the converter gets damaged. The datasheet of a fuse will provide the Total Clearing $I^2t$, which indicates the thermal energy through the fuse until the current is completely interrupted. Using figure \ref{fig:cap_current}, the $I^2t$ of the short-circuit current and the time at which the $I^2t$ of the fuse is reached can be calculated.

\begin{figure}[htbp]
    \centering
    \includegraphics[width=0.45\textwidth]{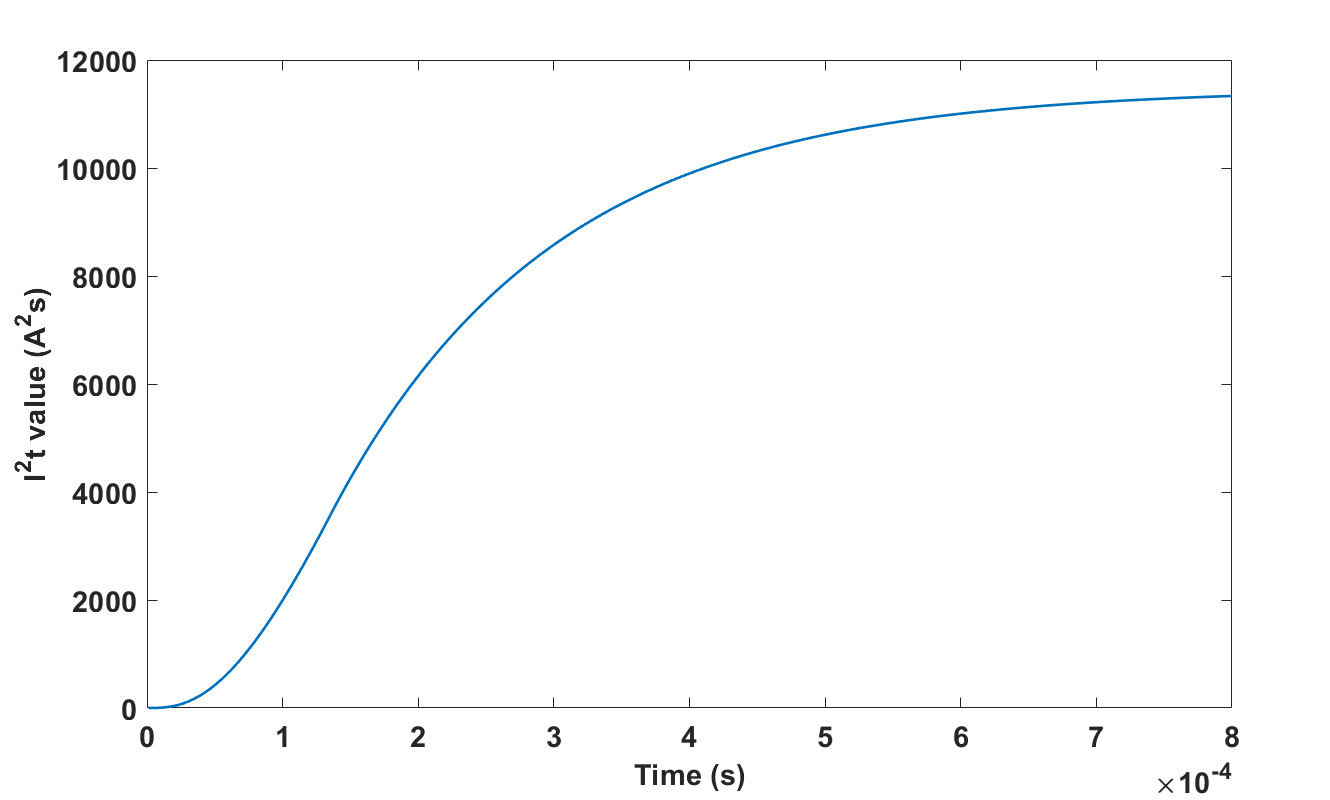}
    \caption{Short-circuit current $I^2t$ curve over time}
    \label{fig:i2t-time}
\end{figure}

If figure \ref{fig:cap_current} is paired with an Eaton Bussmann 170M1790 with an $I^2t$ of 9350, the $I^2t$ of the short-circuit current is equal to that of the fuse at t=0.351ms. The $I^2t$ over time is shown in figure \ref{fig:i2t-time}, and the $I^2t$ value of the fuse should always intercept this curve. Otherwise, it will not clear the fault.

\section{Conclusion}
This paper presents two simulation models in ETAP based on actual vessels that can be adapted for the engineering of future hybrid vessels. For the AC grid simulation, a time domain analysis, short-circuit calculations, and a protection \& coordination basis have been presented, and differences between simulated and calculated results are discussed. For the DC grid simulation, some software limitations were encountered such as missing components and features. Nonetheless, short-circuit calculations and simulations have been presented where possible, and a protection strategy based on $I^2t$ has been discussed.

\section*{Acknowledgment}
The authors thank Alewijnse Netherlands for collaborating on this paper and providing the single-line diagrams of the simulated systems and their parameters.

\printbibliography
\addcontentsline{toc}{chapter}{Bibliography}

\end{document}